\documentstyle[preprint,floats,aps,epsfig]{revtex}
\begin{document}
%\baselinestretch{5}
\tightenlines

\title{Enhancement of $\phi$ mesons in $Pb+Pb$ collisions at 158A GeV/c} 
\vspace{0.1in}
\author{Ben-Hao Sa$^{1,2,5,6}$\footnotemark,
 \footnotetext{E-mail: sabh@iris.ciae.ac.cn} 
 Amand Faessler$^2$,  
  C. Fuchs$^2$, An Tai$^3$, Xiao-Rong Wang\footnotemark,
 Nu Xu$^4$\\ \footnotetext{Present address: Institute of Particle Physics, 
 Huazhong Normal University, Wuhan, 430079 China}
 and E. Zabrodin$^2$} 

\address{
$^1$  China Institute of Atomic Energy, P. O. Box 275 (18), 
      Beijing, 102413 China \\
$^2$  Institute for Theoretical Physics, University of Tuebingen, 
      Auf der Morgenstelle 14, D-72076 Tuebingen Germany \\
$^3$  Department of Physics and Astronomy, University of California, 
      at Los Angeles, Los Angeles, CA 90095 USA \\   
$^4$  Lawrence Berkeley National Laboratory,  
      Berkeley, CA 94720, USA \\
$^5$  CCAST (World Lab.), P. O. Box 8730 Beijing, 100080 China\\
$^6$  Institute of Theoretical Physics, Academia Sinica, Beijing,  
      100080 China
}

\maketitle
\begin{abstract}
Using a hadron-string cascade model LUCIAE, the $\phi$ meson production 
in heavy ion collisions ($Pb+Pb$) and  elementary collisions ($p+p$) 
were studied systematically. Within the framework of the model, the
experimentally measured $\phi$ enhancement in $Pb+Pb$ over $p+p$ 
collisions can be mostly explained by the collective effects in the 
gluon string  emission and the reduction of the $s$-quark suppression.\\  
\noindent{PACS numbers:  25.75.Dw, 24.10.Lx, 24.85.+p, 25.75.Gz}  
\end{abstract}
\vspace{0.1in}

Already suggested in the early eighties \cite{raf1} 
strangeness enhancement is presently considered as one of the 
most promising signatures for the creation of a Quark-Gluon Plasma 
(QGP) phase in relativistic nuclear collisions. At the CERN SPS 
the WA97 Collaboration has measured a clear enhancement of 
multi-strange baryons ($\Lambda , \Xi , \Omega$) in 158A GeV/c 
$Pb+Pb$ collisions relative to $p+Pb$ collisions \cite{wa97}. 
As the mesonic 
counterpart, also an enhancement of the $\phi$ meson production 
in relativistic nuclear collisions was suggested as an 
evidence of the QGP formation in Ref. \cite{shor}, since in the 
environment of a QGP the copious strange and antistrange quarks 
originating from gluon annihilation would be very likely to coalesce 
forming $\phi$ mesons during the hadronization period.
Due to the small cross sections of $\phi$ mesons interacting with 
non-strange hadrons \cite{raf1,shor}, penetrating $\phi$ 
mesons are also messengers   
of the early stage of the colliding system. 
Thus, the $\phi$ meson is not only a promising  
signature for the QGP formation but also a good probe to study the 
reaction dynamics.

Strangeness enhancement in relativistic nucleus-nucleus 
collisions has in the meantime been investigated by various types of 
models, besides LUCIAE \cite{tai1,sa2}. These are: thermal models 
assuming an equilibrated quark      
gluon plasma phase \cite{raf,pet,red}, the non-equilibrium hadron gas 
model with a hadronic strangeness saturation factor \cite{gaz},  
the RQMD \cite{sor} and SFM \cite{ame} models including the  
fusion of overlapping strings, HIJING \cite{wan} and  
HIJING with modifications of the baryon junction exchange mechanism 
\cite{gyu}, UrQMD with a reduction of the constituent quark masses 
\cite{soff} or with a strong color field effect \cite{nu}, 
the diquark breaking model \cite{cap}, and the model of   
strangeness content in nucleon \cite{liu}, etc.  

Recently, NA49 measured the $\phi$ yield, the rapidity and transverse 
mass distributions in $p+p$ and $Pb+Pb$ collisions at 158A GeV/c 
\cite{na49}. The model 
studies on the $\phi$ meson enhancement in relativistic nucleus-
nucleus collisions are rare and to our 
knowledges there exists up to now no theoretical description 
of the full set of NA49 data on the $\phi$ production. 
In this letter we use a hadron and string cascade model,  
LUCIAE \cite{sa}, in order to investigate their data and the 
enhancement mechanism especially. We have successfully used 
LUCIAE to study the enhanced production of multi-strange  
baryons ($\Lambda , \Xi , \Omega$) and determined 
the model parameters related to production of strang 
particles \cite{tai1,sa2}. Therefore, there is no additional 
free parameter in the present calculations for $\phi$ meson production.

The LUCIAE model is based on FRITIOF \cite{pi}, which is an 
incoherent 
hadron multiple scattering and string fragmentation model. In FRITIOF, 
the nucleus-nucleus collision is depicted simply as a superposition of 
nucleon-nucleon collisions. What characterizes LUCIAE beyond FRITIOF 
are the following features: 
First of all, the rescattering between the participant and spectator  
nucleons and the produced particles from the string fragmentation processes  
are generally taken into account \cite{sa1}. However, as proposed 
in \cite{raf1,shor} we, in this work, assume that the final state 
interaction plays no significant role for the $\phi$ production. 
Thus, effects of the final state interactions on 
the  $\phi$ meson production and propagation are neglected. 
Secondly, the collective effect  
in the gluon emission of strings is considered by so-called 
firecracker model \cite{tai}. In relativistic heavy ion collisions the string 
density can be quite high such that some strings might form a collective state. 
Such a string state may emit gluons using its 
larger common energy density. Thirdly, a phenomenological    
mechanism for the reduction of the $s$ quark suppression in the string 
fragmentation process \cite{tai1} is introduced. It is well known that the 
$s$ quark suppression factor (the suppression of $s$ quark pair production 
with respect to $u$ or $d$ pair production in the string fragmentation), 
i. e. the parameter $parj(2)$ in JETSET 
which runs together with FRITIOF and deals with the string 
fragmentation, is not a constant but energy dependent in hadron-hadron 
collisions \cite{tai1,gaz}. In $p+A$ and $A+A$ collisions $parj(2)$  
depends even on the size and centrality of collision system as a result of 
mini-jet (gluon) production stemming from the string-string 
interactions. The phenomenological mechanism introduced in \cite{tai1} 
considers all of the above facts via the effective string tension and 
therefore the pertained JETSET parameters. The  extra model 
parameters introduced were fixed by fitting to $p+p$ data \cite{tai1}.   

\begin{table}[hb]
\begin{tabular}{cccccc}
\multicolumn{6}{c}{Table 1. Average multiplicities of particles in an event}\\
\multicolumn{6}{c}{(momentum 158 GeV/c per nucleon)}\\
\hline\hline
        &       &n$_{ch}$ &n$_{\pi}$ &n$_{\phi}$       &n$_{\phi}$/n$_{\pi}$ \\
\hline
 $p+p$  &NA49   &7.2      &2.87$^*$  &0.012$\pm$0.0015 &0.00418$\pm$0.00053 \\
        &       &         &2.61$^@$  &                 &0.00460$\pm$0.00053$^@$\\
        &LUCIAE &7.82     &2.67      &0.0141           &0.00528           \\
\hline
$Pb+Pb$ &NA49   &         &611       &7.6$\pm$1.1      &0.0124$\pm$0.0018   \\
        &LUCIAE &         &679       &7.89             &0.0116            \\
\hline
\hline
\multicolumn{6}{c}{$^*$taken from Nucl. Phys., B84(1975)269 by NA49}\\
\multicolumn{6}{c}{$^@$after correction for the trigger of pions}\\
\end{tabular}
\end{table}

\begin{table}[hb]
\begin{tabular}{cccc}
\multicolumn{4}{c}{Table 2. Average multiplicities of particles in an event}\\
\multicolumn{4}{c}{of central $Pb+Pb$ collisions at 158A GeV/c}\\
\hline\hline
              &n$_{\pi}$ &n$_{\phi}$       &n$_{\phi}$/n$_{\pi}$ \\
\hline
 LUCIAE       &679       &7.89             &0.0116  \\
 w/o 's'$^*$  &687       &6.28             &0.00914 \\
 w/o 'f'$^@$  &679       &4.29             &0.00632 \\
 w/o s and f  &643       &5.48             &0.00852 \\
\hline
\hline
\multicolumn{4}{c}{$^*$without reduction of s quark suppression}\\
\multicolumn{4}{c}{$^@$without firecracker}\\
\end{tabular}
\end{table}
    
\begin{table}[hb]
\begin{tabular}{ccccc}
\multicolumn{5}{c}{Table 3. The values of four JETSET parameters in}\\
\multicolumn{5}{c}{central $Pb+Pb$ collisions at 158A GeV/c}\\
\hline\hline
            &$parj(1)$   &$parj(2)$   &$parj(3)$   &$parj(21)$ \\
\hline
LUCIAE      &0.116       &0.313       &0.409       &0.373\\
w/o s       &0.100       &0.300       &0.400       &0.320\\
w/o f       &0.0497      &0.215       &0.313       &0.318\\
w/o s and f &0.100       &0.300       &0.400       &0.320\\
\hline
\hline
\end{tabular}
\end{table}

In Table. 1 the LUCIAE results for the $\phi$ meson yield
and the average multiplicities of $\pi^+$ and $\pi^-$, etc. are  
compared to the NA49 data. It should be mentioned here that the pion 
multiplicities were quoted by  NA49 from \cite{ros} where the  
experiment triggers on the total inelastic reaction cross section 
while only 91\% of this cross section was measured in the NA49    
experiment. Thus, a correction must be made 
which is referred to as 'after correction' in Table. 1 . The experimental 
result for the $\phi$-enhancement factor ($\displaystyle \frac{<\phi>/<\pi>
(Pb+Pb \hspace{0.2cm}  central)}{<\phi>/<\pi>(p+p \hspace{0.2cm} 
inelastic)}$) \cite{na49} in $Pb+Pb$ relative to  
$p+p$ after correction is 2.7$\pm$0.7 and the corresponding \\LUCIAE
result is 2.2 . The transverse mass distributions and the 
rapidity distributions of $\phi$ mesons in $p+p$ and $Pb+Pb$ collisions 
at momentum 158 GeV/c per nucleon are compared in Fig. 1 . 
Fitting rapidity distributions obtained from LUCIAE with a 
Gaussian, $f(y)=c\times\exp{(-(y-y_{cm})^2/2/\sigma^2)}$, one obtains 
$\sigma$=0.967 ($p+p$) and 1.05 ($Pb+Pb$), which should be compared with 
the NA49 results of 0.89$\pm$0.06 and 1.22$\pm$0.16, respectively. Since the
inverse slope parameter T extracted from the transverse mass distributions
is very sensitive to the details of the fitting procedure we fit the
highest four $m_t$ data points both for the NA49 and LUCIAE transverse
mass distributions of $Pb+Pb$ collisions with an exponential of form 
$f(m_t)=c\times\exp{(-m_t/T)}$. We obtain then $T_{NA49}$=289 MeV  
and $T_{LUCIAE}$=212 MeV. For $p+p$, if one fits the highest three 
$m_t$ data points both for the NA49 and LUCIAE results one obtains 
nearly the same inverse slope parameter T=189 MeV. To further improve the  
agreement between the data and LUCIAE results one might need to invoke  
the intrinsic transverse momentum broadening in string fragmentation 
\cite{lar} provided the rescattering of $\phi$ meson is not important. 
However, one sees from Table. 1 and Fig. 1 that employing the mechanisms 
of collective string effects in the gluon emission and the reduction  
of the $s$ quark suppression in the string fragmentation 
process, LUCIAE is able to describe, to certain extent, both the data of 
$p+p$ and $Pb+Pb$ collisions consistently.
  
\begin{figure}[ht]
\centerline{\hspace{-0.5in}
\epsfig{file=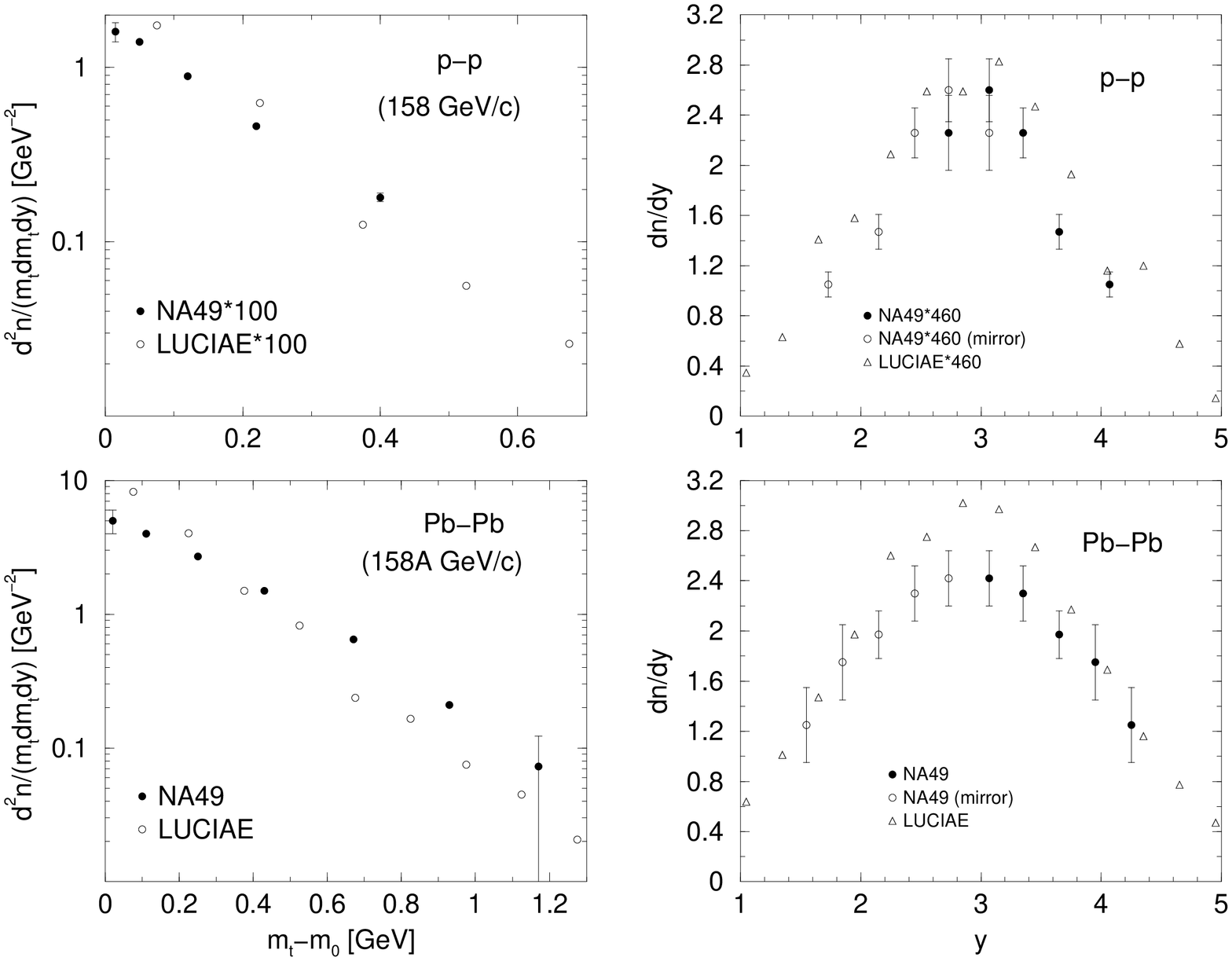,width=4.5in,height=4.5in,angle=-90}}
\vspace{0.2in}
\caption{Transverse mass distributions (left side) and rapidity 
 distributions (right side, 3.0$<y<$3.8 for $p+p$ and 2.9$<y<$4.4 for 
 $Pb+Pb$) of $\phi$ mesons in $p+p$ 
 (upper panels) and $Pb+Pb$ (lower panels) collisions at 
 158A GeV/c}
\label{mty}
\end{figure}

The roles of the mechanisms of the collective effect in the gluon  
emission of strings and the reduction of the $s$ quark suppression in   
string fragmentation in the $\phi$ enhancement are investigated in  
Table. 2. In order to understand the results shown in Table. 2 the   
JETSET parameters relevant to the effective string tension    
are given in Table. 3 . These are the parameters $parj(1)$, $parj(3)$ 
and $parj(21)$, besides $parj(2)$.     
$parj(1)$ stands for the suppression of diquark-antidiquark pair 
production compared to the quark-antiquark pair production in the string 
fragmentation, $parj(2)$ is the suppression of $s$ quark pair production
with respect to $u$ or $d$ pair production,
$parj(3)$ refers to the extra suppression of strange diquark  
production compared to the normal suppression of the strange quark pair. 
Finally $parj(21)$ is the width of the Gaussian transverse momentum 
distribution of $q\bar{q}$ pairs in the string fragmentation. 

The values of these parameters given in the second or fourth line of   
Table. 3 are the default values in JETSET. The mechanism of the reduction 
of $s$ quark  
suppression is considered in a phenomenological way where the effective  
string tension is linked mainly to the transverse momentum of the hardest 
gluon on a string  \cite{tai1}. In  case without the 
firecracker mechanism, but with the reduction of $s$ quark suppression, 
the transverse momentum of gluons on a string is small and thus the values 
of the JETSET   
parameters are smaller than the default values correspondingly   
(cf. lines three and four of Table. 3). That is the reason 
why the $\phi$ meson yield in case without firecracker but with the  
reduction of $s$ quark suppression is even lower than in the case 
without both, firecracker and the reduction of $s$ quark suppression (cf. 
lines three and four of Table. 2). A note is in order here, LUCIAE calculations
with the default JETSET parameters which are determined using $e^+e^-$ data
overestimate production of strange particles in $p+p$ collisions \cite{tai1}, 
which is the very reason that we proposed a phenomenological mechanism to 
investigate how the string tension varies as a function of collision energy in  
$p+p$ collisions. One can see from Table. 1 and 2 that 
the firecracker model plays the major role and the reduction 
of $s$ quark suppression is  
significant only in combination with the firecracker model.   

It is interesting to compare LUCIAE \cite{tai1} with UrQMD \cite{soff,nu} 
in the mechanism of strangeness enhancement. Both of them start 
from the quantum tunneling probability 
\begin{equation}
\exp{(\frac{-\pi m^2}{\kappa})}\exp{(\frac{-\pi p_t^2}{\kappa})}
\end{equation}  
for the production of $q\bar{q}$ pair with the quark mass $m$ and the 
transverse momentum $p_t$ from a string of string tension $\kappa$ 
\cite{sch}. Thus, the suppression factor of the $s\bar{s}$ pair production 
with respect to $u$ or $d$ pair, for instance, can be expressed as
\begin{equation}
parj(2)=\exp{(\frac{-\pi(m_s^2-m_u^2)}{\kappa})}.
\end{equation}
In \cite{soff,nu} it was then argued that in the relativistic $A+A$ 
collisions the string tension should be three times larger than that 
in $p+p$ collisions at the same energy due to the higher string density. 
The increase of the string tension is further attributed to the reduced 
quark mass stemming possibly from a transition of chiral  
restoration \cite{soff}. On the other hand, in \cite{tai1} an effective 
string tension was introduced and is phenomenologically related to 
the multigluon string in comparison with the pure $q\bar{q}$ string.  
Consequently the effective string tension and the pertained JETSET 
parameters are increasing with the energy, the size and centrality 
of collision system. Therefore, strangeness production in 
relativistic $p+p$, $p+A$ and $A+A$ collisions 
might be investigated consistently within a hadron-string model without 
introducing the QGP formation explicitly. An interesting issue arised 
here is worthy to be studied further. 
We also plan to improve the agreement
between the experimental rapidity and transverse mass distributions 
and the LUCIAE results via transverse excitation of string and the 
intrinsic transverse momentum  
broadening in string fragmentation. The investigation for the role   
played by the hard and semi-hard processes, such as $gg\rightarrow    
s\bar{s}$ and $q\bar{q}\rightarrow s\bar{s}$ on strangeness enhancement
is needed as well. 

In summary, the experimentally found $\phi$ enhancement in $Pb+Pb$
relative to $p+p$ collisions is described consistently by the
hadron-string cascade model LUCIAE. In this model, $\phi$ mesons are
exclusively produced from string fragmentation processes without any
further rescattering interactions. However, LUCIAE has employed the
mechanisms of the string collective effect (firecracker model) in the
gluon emission and of the reduction of $s$-quark suppression in the
string fragmentation. This implies that, at the CERN SPS energy, the
$\phi$-mesons are mostly produced in primordial collisions and final
state interactions at the hadronic stage do not play a significant
role.

Finally, the financial supports from NSFC in China, DFG in Germany,   
and DOE in USA are acknowledged.

\end{document}